\documentclass[11pt]{article}
\usepackage{moriond,epsfig}

\def\Journal#1#2#3#4{{#1} {\bf #2}, #3 (#4)}

\def\NPB{{\em Nucl. Phys.} B}
\def\PLB{{\em Phys. Lett.}  B}

\def\PRD{{\em Phys. Rev.} D}

\def\JHEP{{\em JHEP.}}

\begin{document}

\begin{flushright}
LAPTH-Conf-1243/08\\
Avril 2008
\end{flushright}

\vspace*{4cm}
\title{THE SIX-PHOTON AMPLITUDES}

\author{C. Bernicot }

\address{LAPTH, Universit\'e de Savoie, CNRS\\
B.P. 110, F-74941 Annecy-le-Vieux Cedex, France.}

\maketitle\abstracts{Thanks to the absence of tree order, the
six-photon processes is a good laboratory to study multi-leg
one-loop diagrams. Particularly, there are enough on-shell
external legs to observe a special Landau singularity: the double
parton scattering.}

\section{Introduction}

\subsection{Motivations}

At LHC, we hope to discover new physics by the collisions between
two protons. The partonic processes constitute a background, which
is mandatory to know, if we want to observe new particles. In QCD,
the coupling constant depends on an unphysical energy scale and to
reduce this dependency, we have to increase the order of the
expansion. So, new efficient methods, based on unitarity, have
been developed \cite{BrittoCUT,MastroliaCUT}, for the NLO (Next to
Leading Order) calculation. As the six-photon amplitudes have no
rational terms, and no divergences therefore they are a good
laboratory to apply those methods to multi-leg one-loop diagrams.

\subsection{Difficulties of NLO calculation}

The amplitude of a six-photon diagram is the product of two terms,
a tensor with the polarisation vectors of the external photons and
a tensor integral. The first difficulty is to find a clever
formulation of the polarisation vectors to simplify the expression
and to obtain a compact result, and the second is to reduce
efficiently the tensors integrals. The two solutions are the
spinor formalism with helicity amplitudes, described in
\cite{spinor} and the efficient reduction thanks to unitary-cuts
\cite{BrittoCUT,MastroliaCUT}.

\section{Results and Plots}

In the past, three teams have calculated the six photons
amplitudes analytically or numerically in QED
\cite{Nagy,binoth:2007,Papadopoulos:6photons}. I obtain very
compact expressions for all the six-photon helicity amplitudes in
QED, scalar QED and supersymmetric $\textrm{QED}^{{\cal N} = 1}$
\cite{Bernicot:6photons}. Each amplitude is a linear combination
of four-point scalar integrals in $n+2$ dimensions and three point
three-external-mass scalar integrals in $n$ dimensions.

\hfil

\begin{figure}[httb!]
\centering
\includegraphics[width=10cm]{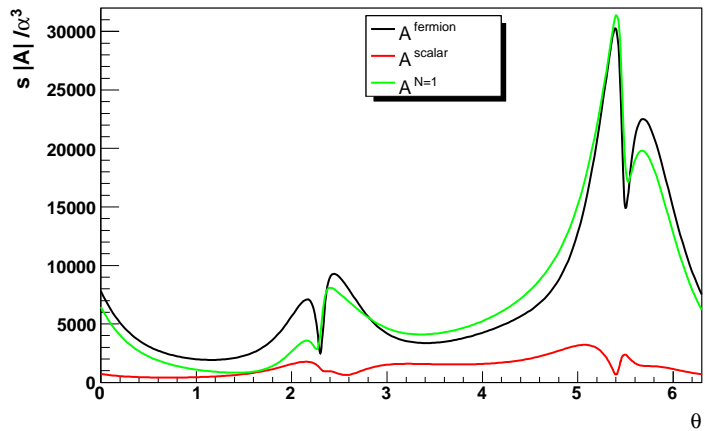}
\caption{NMHV helicity amplitude of the six-photon process
\label{fig:NMHV}}
\end{figure}

\begin{figure}[httb!]
\centering
\includegraphics[width=9cm]{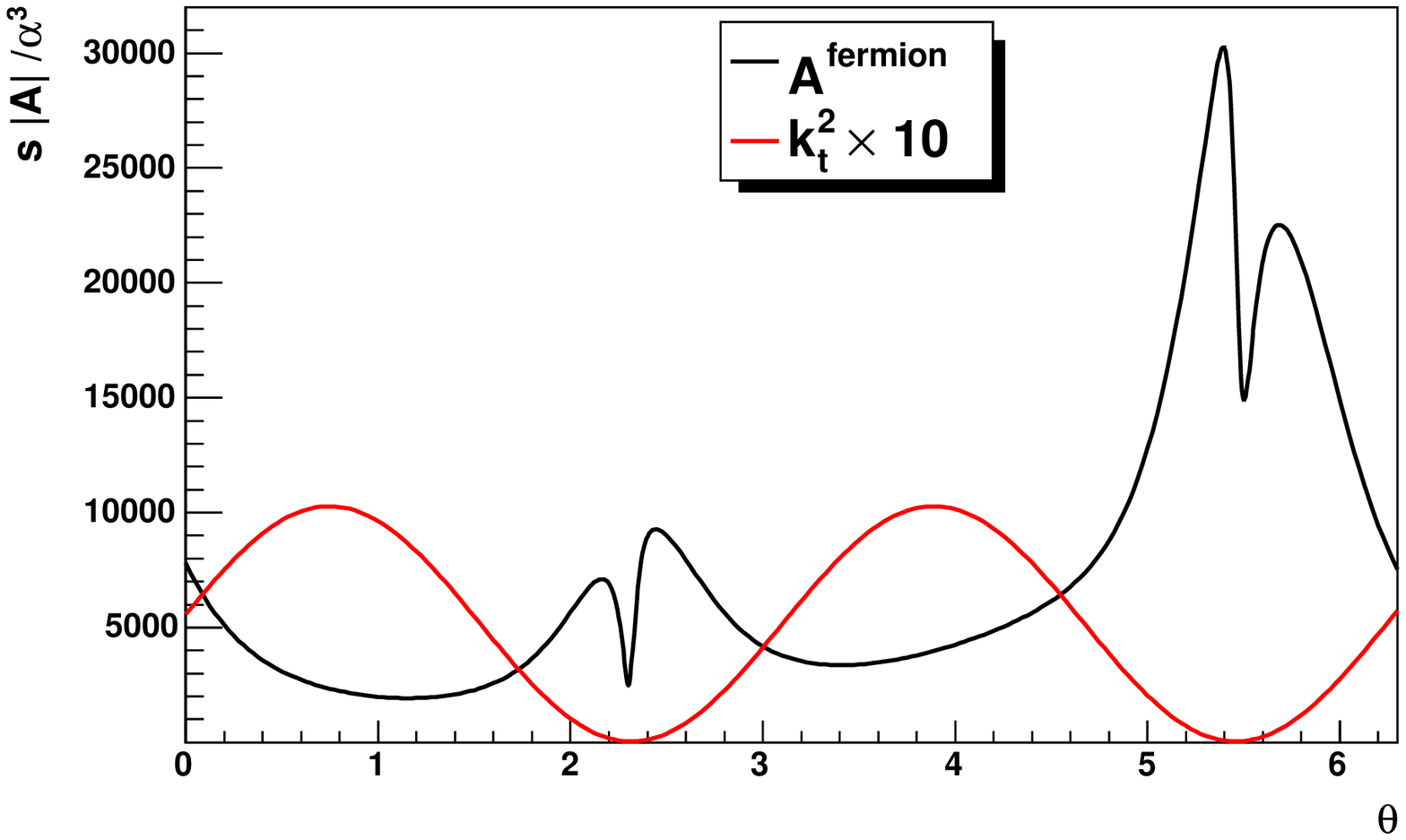}
\includegraphics[width=5cm]{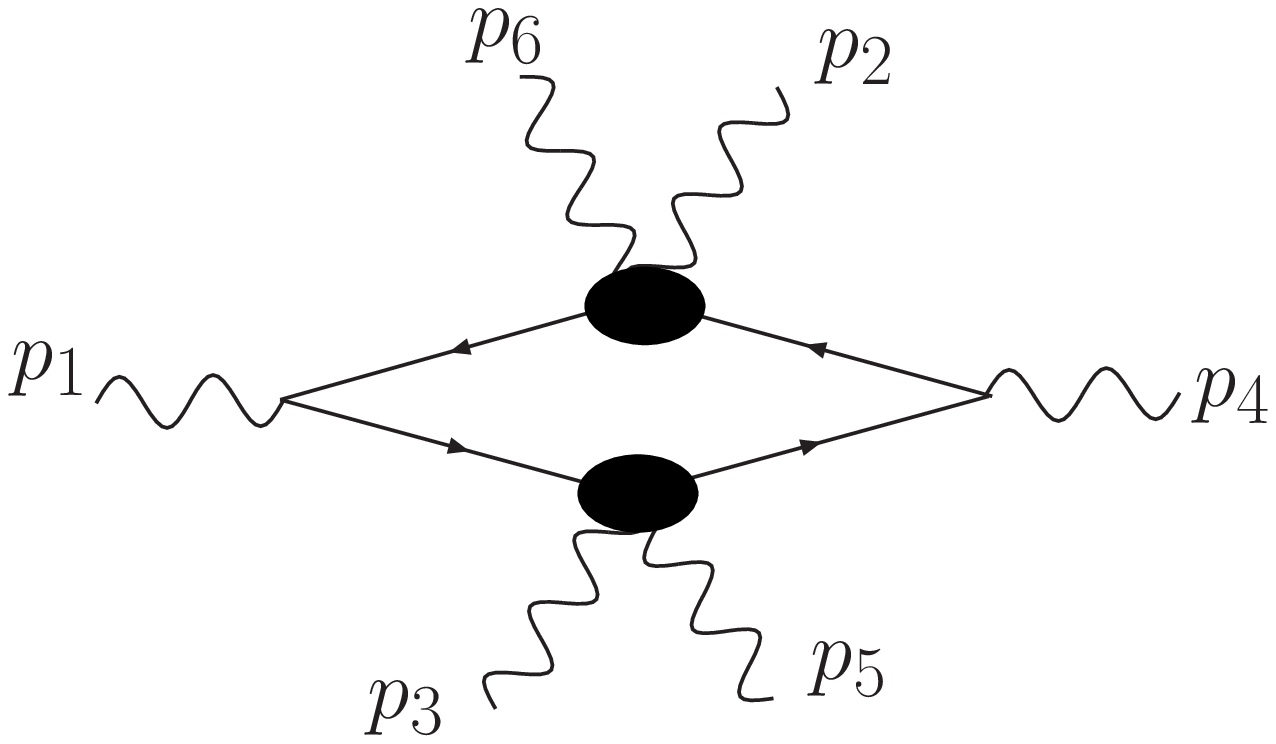}
\caption{Localisation and kinematical configuration of double
partons scattering \label{fig:DPS}}
\end{figure}

Let us plot the amplitudes in the Nagy-Soper kinematical
configuration \cite{Nagy}. The photons 1 and 4 constitute the
initial state along the z-axis whereas the photons 2, 3, 5 and 6
the final state. In the center of mass frame of the initial state,
 we put the final state at the phase space point:
\begin{equation}
    \parbox{3.5cm}{\includegraphics[width=3.5cm]{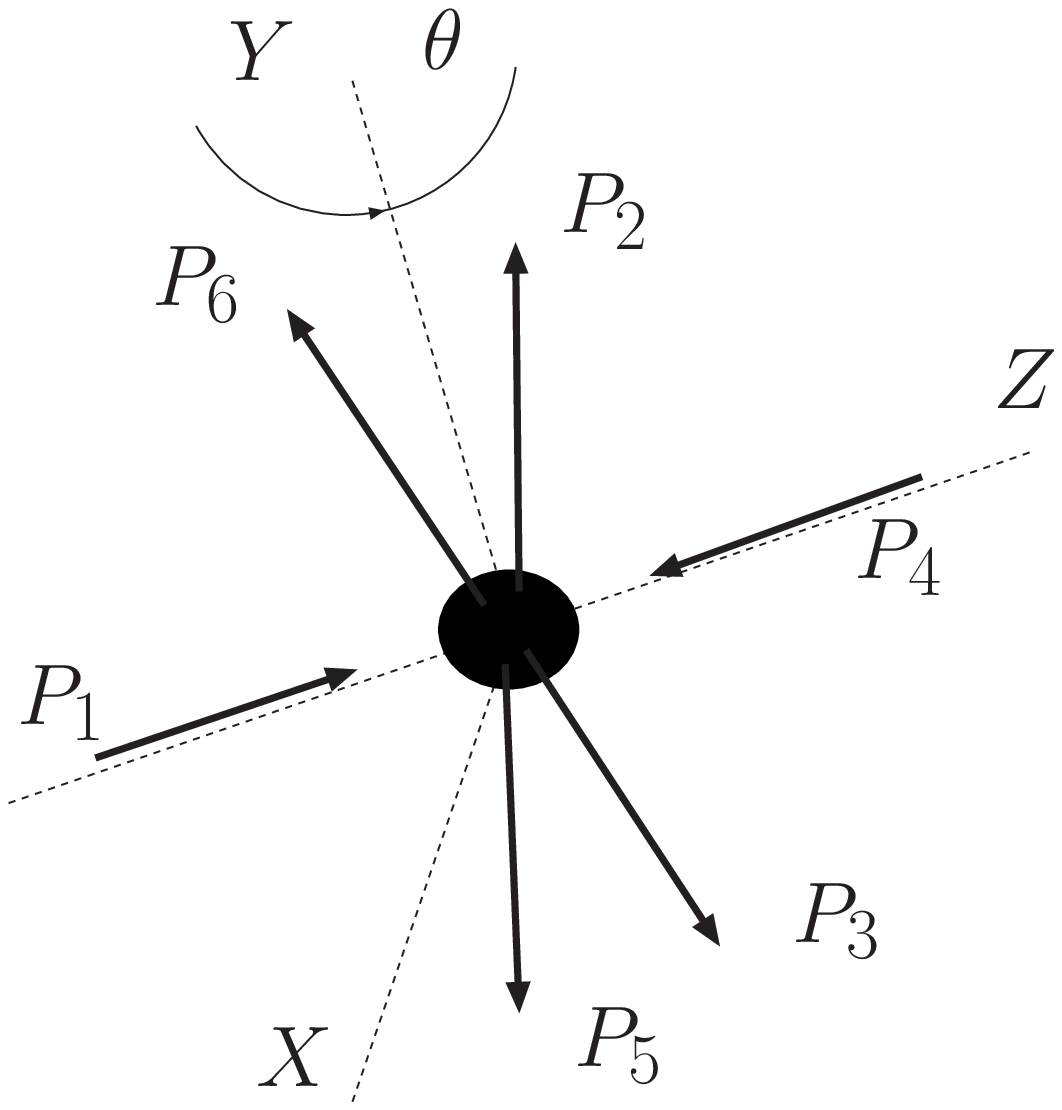}}
    \quad \quad \quad \left\{ \begin{array}{l}
    \overrightarrow{p_{2}} = (-33.5,-15.9,-25.0) \\
    \overrightarrow{p_{3}} = (11.0,13.2,22.0) \\
    \overrightarrow{p_{5}} = (12.5,-15.3,-0.3) \\
    \overrightarrow{p_{6}} = (10.0,18.0,3.3)
    \end{array} \right. \nonumber
\end{equation}
\noindent New kinematical configuration are generated, by rotating
the final state by an angle $\theta$ about the y-axis,
perpendicular to the z-axis. In the figure $\ref{fig:NMHV}$, we
plot the NMHV (Next to Maximal Helicity Violating) amplitudes
versus $\theta$.
\noindent Two peaks appear for each amplitude at the angles
$\theta_{1} \simeq 2.32$ and $\theta_{2} \simeq \pi + 2.32 \sim
5.46$. To understand the origins of these peaks, we split the
final state in two photon pairs $(3,5)$ and $(2,6)$. We note
$k_{t}$ the transverse momentum of each photon pair and we plot
its value versus the angle $\theta$ on the left-graph of the
figure $\ref{fig:DPS}$. The peaks occurs exactly at the points
where $k_{t}$ is the smallest : it is the signature of double
parton scattering. It is a special kinematical configuration,
corresponding to a Landau singularity.

\section{Landau singularity}

Physically, Landau singularities correspond to a "resonance" of
the virtual loop-particle with a physical process. For the
six-photon amplitudes, this physical process is represented by
diagrams on the figure $\ref{fig:DPS}$. The two ingoing photons 1
and 4 split each into a fermion anti-fermion collinear pair, then
each fermion scatters with the anti-fermion to give a photon pair
with no transverse momentum.

\hfil

In a one-loop diagram, a Landau singularity is defined by finite
points in the phase space, where the integrand of the loop is not
analytic. But even if, locally the denominator is zero for
example, the integral may be finite. We want to know if there are
some divergences in the special case of this processes.

\hfil

We reach the singularity when the transverse momentum of each pair
of photons is equal to zero. With the Nagy-Soper kinematical
configuration, we cannot reach it, so we modify it. As we rotate
around the y-axis, we add or subtract a y-momentum $\Delta^{y}$
for each final photons, to keep on the energy-momentum
conservation :
\begin{equation}
\left\{ \begin{array}{l}
    \overrightarrow{k_{3}} = (33.5,15.9-\Delta^{y},25.0) \\
    \overrightarrow{k_{4}} = (-12,5,15.3+\Delta^{y},0.3) \\
    \overrightarrow{k_{5}} = (-10.0,-18.0+\Delta^{y},-3.3) \\
    \overrightarrow{k_{6}} = (-11.0,-13.2-\Delta^{y},-22.0)
\end{array} \right.
\end{equation}
\noindent $\Delta^{y}$ acts as a regulator and the singularity is
reached at $\Delta^{y} = 1.05$. Let us plot the QED amplitude
around the singularity for several values of this regulator:
\begin{figure}[httb!]
\centering
\includegraphics[width=9cm]{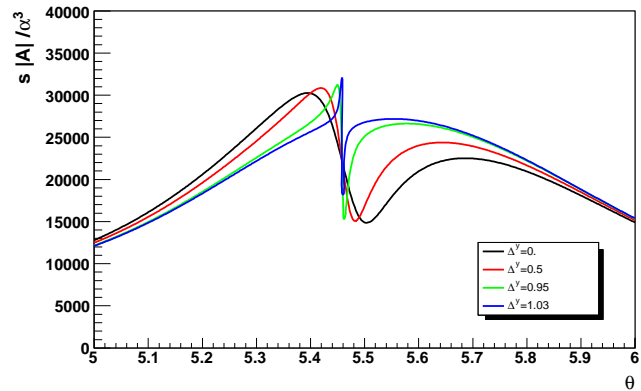}
\caption{Around the double partons scattering
\label{fig:aroundDPS}}
\end{figure}
\noindent The amplitude behaves as a wave around the singularity.
It is larger for $\Delta^y=0$ and disappears completely for
$\Delta^y=1.05$. The closer $\Delta^y$ is from 1.05, the more
squeezed the structure support is. There is no divergence because
the numerator of the six-photon amplitudes vanish at the Landau
singularity fast enough to regularize it. More explanations are
given in \cite{Houches}.

\subsection{Summary}

The six-photon amplitude is a good laboratory to study one-loop
multi-legs diagrams, particularly the "analycity" of the integrand
of the loop. The non-analytic phase space points, called Landau
singularities, let traces when plotting the amplitude (the double
parton scattering). Fortunately, the structure of QED regularize
them.

\end{document}